\documentclass[aps,prd,twocolumn,superscriptaddress,showpacs,showkeys]{revtex4-1}  
\usepackage{graphicx,color} 
\usepackage{xcolor}
\usepackage{rotating}
\usepackage{amssymb}
\usepackage{amsfonts}
\usepackage{amsmath}
\usepackage{xspace}
\usepackage{mathtools} 
\usepackage{verbatim}

\usepackage{listings}
\makeindex

\usepackage[thinlines]{easytable}

\usepackage[super]{nth}
\usepackage{braket}

\newcommand{\be}{\begin{equation}}
\newcommand{\ee}{\end{equation}}
\newcommand{\bea}{\begin{eqnarray}}
\newcommand{\eea}{\end{eqnarray}}

\newcommand\ie{\mbox{\textit{i.\,e.}}\xspace}

\newcommand\eg{\mbox{e.\,g.}\xspace}

\newcommand\D{\mathrm{d}}

\newcommand\hil{\mathcal{H}}
\newcommand{\var}[1]{\left(\sigma^2_{#1}\right)}

\begin{document}

\title{Gravitationally induced uncertainty relations in curved backgrounds}




\author{Luciano Petruzziello}
\email{lupetruzziello@unisa.it}
\affiliation{Dipartimento di Ingegneria Industriale, Universit\'a degli Studi di Salerno, Via Giovanni Paolo II 132, I-84084 Fisciano (SA), Italy.}
\affiliation{INFN, Sezione di Napoli, Gruppo collegato di Salerno, Italy.}

\author{Fabian Wagner}
\email[]{fabian.wagner@usz.edu.pl}
\affiliation{Institute of Physics, University of Szczecin, Wielkopolska 15, 70-451 Szczecin, Poland}
\date{\today}
\begin{abstract}
This paper aims at investigating the influence of space-time curvature on the uncertainty relation. In particular, relying on previous findings, we assume the quantum wave function to be confined to a geodesic ball on a given space-like hypersurface whose radius is a measure of the position uncertainty. On the other hand, we concurrently work out a viable physical definition of the momentum operator and its standard deviation in the non-relativistic limit of the 3+1 formalism. Finally, we evaluate the uncertainty relation which to second order depends on the Ricci scalar of the effective 3-metric and the corresponding covariant derivative of the shift vector. For the sake of illustration, we apply our general result to a number of examples arising in the context of both general relativity and extended theories of gravity.
\end{abstract}


\maketitle

\section{Introduction}
\label{intro}

Heisenberg's celebrated uncertainty principle in its fundamental form does not take into account effects which are expected to arise when gravity is being accounted for. In fact, one of the low-energy implications associated with several candidate theories of quantum gravity predicts a correction to the aforementioned relation going by the name Generalized Uncertainty Principle (GUP) \cite{Amati87,Kempf95,Maggiore93,Adler01,GUPReview,AdlerDuality}. After seminal efforts have been devoted to this topic, the investigation revolving around the generalization of the fundamental quantum mechanical uncertainty has developed remarkably. As a matter of fact, the consequences entailed by the GUP deeply affects several aspects of black hole thermodynamics \cite{Sparsity,noncom,noncom2,noncom3,bis,ter,ter2,quar,ses,ong,set,ot,nov,dieci}, quantum field theory \cite{noui,frassino,qft,qft2,qft3,qft4,czech,czech2,pikovski,plenio,plenio2} and also quantum information \cite{dasprl,dasdecoherence,decoherence}. 

As usually assumed, the GUP takes into account the gravitational uncertainty of position in connection with the existence of a minimal fundamental length scale in physics. However, it may not be the only gravitationally induced change. In particular, the curvature of space-time does exert an influence over quantum mechanical uncertainty relations. This is the regime of the Extended Uncertainty Principle (EUP) \cite{Mignemi2010,Anrade,Mureika,EUPThermod,Bambi2008}, whose purpose lies in the investigation of the uncertainty related to the background space-time. 

Recently, a derivation of the EUP was performed in \cite{Schuermann2009,Schuermann2018}. It reflects the influence of spatial curvature on quantum-mechanics on 3-dimensional spacelike hypersurfaces of space-time. The method was applied to homogeneous and isotropic geometries of constant curvature $K$ and the corresponding EUP was calculated. This formalism was further developed in collaboration with one of the authors to accommodate for horizons \cite{RF2019} and arbitrary 3-dimensional manifolds, thus yielding an asymptotic extended uncertainty principle \cite{RF2020}. In short, the quantum mechanical wave function of interest is defined over a curved three-dimensional manifold and its domain constrained to a geodesic ball. Then, the radius of this ball serves as a diffeomorphism-invariant measure of position uncertainty. By computing the standard deviation of the momentum operator, one finds that there is a generic minimum which depends on the radius of the geodesic ball and on curvature invariants derived from the effective background. This represents a mathematically rigorous as well as physically motivated formulation of the EUP.

This paper goes further inasmuch it aims to provide the uncertainty relation experienced by a non-relativistic particle in a curved 4-dimensional background. To comply with this goal, we obtain the effective Lagrangian according to which the particle dynamics evolves. As is well-known in the literature \cite{gravitomagnetism}, the gravitational field splits up into a three-dimensional background metric, a gravitomagnetic vector field and a gravitoelectric scalar field. Due to the resulting gauge invariance which is not shared by the conjugated momentum, the physical momentum includes a contribution from the vector field. Adding these ingredients to the picture outlined above and taking into account subtleties of the quantization as first pointed out in \cite{DeWitt:1952js}, we finally achieve the desired result. This makes it possible to compare the impact of distinct theories of gravity (which describe different backgrounds) on the uncertainty relation.

The paper is organized as follows: in section \ref{sec: effham} we derive the effective Hamiltonian of a non-relativistic particle on a curved 4D background. This theory is quantized in section \ref{sec:QMtreat}, where we define the momentum uncertainty and the relevant operators whose expectation values appear therein and compute the curvature induced corrections to the uncertainty relation. The corresponding results are applied to several relevant space-times in section \ref{sec:app} to be subsequently discussed in section \ref{sec:disc}.

\section{Derivation of the effective Hamiltonian\label{sec: effham}}

In this section, we will derive the effective dynamics of a non-relativistic particle on a curved 4-dimensional background. To this aim, note that the action of a massive relativistic particle subjected to a curved geometry can be written as
\begin{align}
S	&=-m\int \D s\\
	&=-m\int \sqrt{-g_{\mu\nu}(x)\dot{x}^\mu\dot{x}^\nu} \D \tau\label{relaction}
\end{align}
with the background metric $g_{\mu\nu}$ (the Greek indices stand for space-time coordinates while Latin ones denote the spatial part), the four-velocity $\dot{x}^\mu=\D x^\mu/\D\tau$ ($\tau$ denotes the proper time along the curve) and the mass of the particle $m.$ Thus, the corresponding Lagrangian reads
\begin{equation}
L=-m\sqrt{-g_{\mu\nu}(x)\dot{x}^\mu \dot{x}^\nu}.\label{lagrangian2}
\end{equation}
Note that the action \eqref{relaction} is invariant under temporal reparametrisations $\tau'=f(\tau)$ for any sufficiently well-behaved function $f.$

After some algebra, the Lagrangian can be recast as
\begin{align}
L=  &-m\Big[\left(-g_{00}+g_{0i}g_{0j}h^{ij}\right)(\dot{x}^0)^2\nonumber\\
    &-\left(\dot{x}^0g_{0k}h^{ik}+\dot{x}^i\right)\left(\dot{x}^0g_{0l}h^{jl}+\dot{x}^j\right)g_{ij}\Big]^{1/2}
\end{align}
with $h^{ik}g_{kj}\equiv\delta^i_j.$ Thus, $h^{ij}$ is the inverse of the induced metric on hypersurfaces of constant $x^0.$ The Lagrangian can be further simplified by introducing the background field quantities
\begin{align}
    N=&\sqrt{-g_{00}+g_{0i}g_{0j}h^{ij}}\\
    N^i=&g_{0j}h^{ij}
\end{align}
which are readily identified as the lapse function and the shift vector in the 3+1 formalism \cite{ADM}. According to this approach to curved Lorentzian manifolds, any metric can be written as \cite{eric}
\begin{equation}
    \D s^2=-N^2\left(\D x^0\right)^2+h_{ij}(N^i\D x^0+\D x^i)(N^j\D x^0+\D x^j)
\end{equation}
with $h_{ij}\equiv g_{ij}.$ Thence, the Lagrangian reads
\begin{equation}
    L=-m\sqrt{N^2\left(\dot{x}^0\right)^2-\left(\dot{x}^0N^i+\dot{x}^i\right)\left(\dot{x}^0N^j+\dot{x}^j\right)h_{ij}}
\end{equation}
which under the assumption that $\dot{x}^0>0$ (\ie that coordinate time is moving in the same direction as the particle proper time) can be written as
\begin{align}
     L=&-mN\dot{x}^0\sqrt{1-\frac{\left(\dot{x}^0N^i+\dot{x}^i\right)\left(\dot{x}^0N^j+\dot{x}^j\right)h_{ij}}{N^2\left(\dot{x}^0\right)^2}}\\
      \equiv&-mN\dot{x}^0\sqrt{1-\epsilon}
\end{align}
where the last equality defines $\epsilon.$ In terms of the conjugate momenta $P_\mu=mg_{\mu\nu}\dot{x}^\nu/\sqrt{-g_{\mu\nu}\dot{x}^\mu\dot{x}^\nu}$, we find $\epsilon/(1-\epsilon)=h^{ij}P_iP_j/m^2$, which means that the non-relativistic limit corresponds to $\epsilon\ll 1.$ Therefore, we can expand to obtain the effective non-relativistic Lagrangian
\begin{equation}
    L_{NR}=\frac{m}{2\dot{x}^0}\left(\dot{x}^0N^i+\dot{x}^i\right)\left(\dot{x}^0N^j+\dot{x}^j\right)G_{ij}-mN\dot{x}^0
\end{equation}
with the effective 3-metric $G_{ij}=h_{ij}/N.$ From this point onwards, this metric will be used to lower and raise indices and as the background for differential geometric quantities.

The effective non-relativistic action $S_{NR}=\int L_{NR}\D\tau$ still harbours the time reparametrization invariance alluded to above. For simplicity, we can fix the gauge by choosing $x^0=\tau$ to obtain the gauge-fixed non-relativistic Lagrangian
\begin{equation}
    L_{NR}=\frac{m}{2}\left(N^i+\dot{x}^i\right)\left(N^j+\dot{x}^j\right)G_{ij}-mN.
\end{equation}
A closer look at the Lagrangian tells us that it is of the form
\begin{equation}
    L_{NR}=\frac{m}{2}\dot{x}^i\dot{x}^jG_{ij}+m\dot{x}^iA_i-m\phi
\end{equation}
with $A_i=N^jG_{ij}$ and $\phi=N-N^iN^jG_{ij}/2.$ This is clearly reminiscent of the Lagrangian describing a charged non-relativistic particle minimally coupled to an electromagnetic gauge one-form $A_\mu=(\phi,A_i)$ where the mass $m$ plays the rôle of the charge.

On the other hand, the Lagrangian is additionally invariant under the gauge transformation $A\rightarrow A_i+\partial_i f,$ $\phi\rightarrow\phi-\dot{f},$ $G^{ij}\rightarrow G^{ij}$ for any scalar function $f(x^i,t)$ while the canonical momenta
\begin{equation}
    \pi_i=\frac{\partial L_{NR}}{\partial \dot{x}^i}=mG_{ij}\left(\dot{x}^j+N^j\right)
\end{equation}
are not, rendering them unobservable. Therefore, we define the gauge invariant physical momenta as
\begin{equation}
    p_i\equiv\pi_i-mN^jG_{ij}\label{phys}
\end{equation}
in terms of which the Hamiltonian reads
\begin{equation}
    H_{NR}=\frac{1}{2m}p_ip_jG^{ij}+m\phi.\label{effhamiltonian}
\end{equation}
Having found the effective Hamiltonian, we are now able to give a quantum mechanical description of non-relativistic particles in curved space-time.

\section{Quantum Mechanical treatment\label{sec:QMtreat}}

Aiming towards a generalization of the reasoning in \cite{RF2020} to four dimensional backgrounds, we have to find the quantum mechanical counterpart of the theory outlined in the previous section. For this purpose, we first construct an equivalent Hilbert space and introduce the basis in terms of which calculations will be performed. As we pointed out above, the effective Hamiltonian \eqref{effhamiltonian} entails a certain amount of gauge freedom which makes it necessary to define the gauge invariant (\ie physical) momenta $p_i.$ Moreover, we obtain the quantum operators representing their co- and contravariant versions and define the standard deviation accordingly. To conclude the section, we derive the flat space uncertainty relation and its higher-order corrections.

\subsection{Hilbert space and used basis} 

In order to construct the Hilbert space, we have to find an appropriate measure defining the scalar product $\braket{\psi|\phi}=\int\D\mu\psi^*\phi.$ The problem at hand takes place in a 3-dimensional curved background defined by the canonical connection with respect to $G_{ij}$ and offers the gauge freedom mentioned above. These facts should be reflected by the measure making it gauge and diffeomorphism invariant. The only possible differential satisfying these constraints reads $\D\mu=\sqrt{G}\D^3x$ where $G=\text{det}G_{ij}.$

Furthermore, we confine the quantum states to a geodesic ball $B_\rho$ of radius $\rho$ by imposing Dirichlet boundary conditions as in \cite{RF2020}; hence, they reside in the Hilbert space $\mathcal{H}=L^2(B_{\rho}\subseteq {\rm I\!R}^3,\sqrt{G}\D^3 x).$ As the center of the geodesic ball $p_0$ is situated in the background manifold, it represents the expectation value of the position operator while $\rho$ can be interpreted as a measure of position uncertainty.

Since the negative Laplace-Beltrami operator $-\Delta$ is hermitian in $\mathcal{H},$ its eigenstates furnish an orthonormal basis of the Hilbert space. Due to the compactness of the domain of $\mathcal{H}$, the spectrum of said operator is discrete. Thus, there is a countably infinite number of base vectors satisfying the eigenvalue problem
\begin{equation}
(\Delta +\lambda_{nlm})\psi_{nlm}=0\hspace{1cm}\psi_{nlm}|_{\partial B_\rho}=0 \label{evprob}
\end{equation}
where $n,l,m$ stand for the three quantum numbers identifying the state. 

In the case of a flat background, the solutions to the problem \eqref{evprob} can be obtained analytically and read in normalized form
\begin{equation}
    \psi_{nlm}=\sqrt{\frac{2}{\rho^3 j^2_{l+1}(j_{l,n})}}j_l\left(x_{nl}\frac{\sigma}{\rho}\right)Y^m_l(\chi,\gamma)\label{psinlm}
\end{equation}
with the spherical Bessel function $j_l,$ the spherical harmonics $Y^m_l$ and the pure number $j_{l,n}$ which is the nth solution of the equation $j_l(j_{l,n})=0$ or, in other words, the nth zero of the Bessel function $J_{l+1/2}.$ The corresponding eigenvalues are 
\begin{equation}
    \lambda_{nl}=\frac{j_{l,n}^2}{\rho^2}.\label{flatev}
\end{equation}

\subsection{Physical momentum operator}

When promoting positions and momenta to quantum operators, we impose the canonical commutation relations $[\hat{x}^i,\hat{p}_j]=i\hbar\delta^i_j.$ In the position space representation, these are solved by the operators
\begin{equation}
\hat{x}^i\psi=x^i\psi\hspace{1cm}\hat{p}_i\psi=\left(-i\hbar\partial_i+F_i(x)\right)\psi
\end{equation}
where $\psi$ describes a general wave function. The arbitrary form $F_i(x)$ can be found imposing that the momentum operator be hermitian with respect to the Hilbert space measure $\D\mu$, \ie
\begin{align}
\braket{\psi|\hat{p}_i\phi}=\braket{\hat{p}_i\psi|\phi}
\end{align}
and aptly represents the physical momentum in the classical limit.

It can be shown that these criteria are uniquely satisfied by the momentum operator
\begin{align}
\hat{p}_i\psi	=&\left(\hat{\pi}_i-m\hat{N}^j\hat{G}_{ij}\right)\psi\\
					\equiv&-\left[i\hbar\left(\partial_i+\frac{1}{2}\Gamma^j_{ij}\right)+mN^jG_{ij}\right]\psi
\end{align}
where the second equality defines $\hat{\pi}_i$ and we introduced the Christoffel symbol
\begin{align}
\Gamma^k_{ij}=\frac{1}{2}G^{kl}\left(\partial_iG_{jk}+\partial_jG_{ik}-\partial_kG_{ij}\right).
\end{align}
Using this result, there is only one possible hermitian definition of the covariant momentum operator reading
\begin{align}
\hat{p}^i\psi	=&\left(\hat{\pi}^i-m\hat{N}^i\right)\psi\\
						\equiv&-\left[\frac{i\hbar}{2} \left(\left\{G^{ij},\partial_j\right\}+G^{ij}\Gamma^k_{kj}\right)+mN^i\right]\psi\\
						=&\frac{1}{2}\left\{\hat{G}^{ij},\hat{p}_j\right\}\psi
\end{align}
where $\hat{\pi}^i$ is defined according to the second equality.

Finally, the square of the momentum operator is of the form
\begin{equation}
\hat{p}^2=\hat{\pi}^2-2m\hat{p}_{mix}^2+m^2\hat{G}_{ij}\hat{N}^i\hat{N}^j
\end{equation}
where $\hat{\pi}^2\psi=-\hbar^2\Delta\psi$ and $\hat{p}^2_{mix}$ mixes canonical momenta and the shift.

Under the assumptions that it is hermitian and leads to the correct classical limit, this operator reads
\begin{align}
\hat{p}^2_{mix}=	&\frac{1}{2}\left\{\hat{\pi}_i,\hat{N}^i\right\}.
\end{align}
After some elementary algebra, it can be seen that it acts on wave functions as
\begin{equation}
\hat{p}^2_{mix}\psi=-\frac{i\hbar}{2}\left[\nabla_i\left(N^i\right)+2N^i\partial_i\right]\psi\label{pmix}
\end{equation}
with the covariant derivative $\nabla_i$ with respect to the canonical connection of $G_{ij}.$

Note that the first term in \eqref{pmix} is anti-hermitian, and since $\hat{p}^2_{mix}$ is hermitian it cancels out the anti-hermitian part of the second term so that we can rewrite $\hat{p}^2_{mix}$ as 
\begin{align}
    \hat{p}^2_{mix}=\left(-i\hbar N^i\partial_i\right)_H\psi
\end{align}
where the subscript $H$ denotes the hermitian part.

Summing up the outcome of this subsection, the relevant operators act as
\begin{align}
\hat{p}_i\psi		=&\hbar\left[-i\left(\partial_i+\frac{1}{2}\Gamma^j_{ij}\right)-\frac{G_{ij}N^j}{\lambdabar_C}\right]\psi\label{conmom}\\
\hat{p}^i\psi	=&\hbar\left[-\frac{i}{2}\left(\left\{G^{ij},\partial_j\right\}+G^{ij}\Gamma^k_{jk}\right)-\frac{N^i}{\lambdabar_C}\right]\psi\label{covmom}\\
\hat{p}^2\psi	=&\hbar^2\left[-\Delta+\frac{2}{\lambdabar_C}\left(iN^i\partial_i\right)_H+\frac{N^iN^jG_{ij}}{\lambdabar_C^2}\right]\psi\label{squmom}
\end{align} 
with the reduced Compton wavelength $\lambdabar_C=\hbar/m.$

Having figured out the position space representation of the contra-, covariant and squared momentum operators in \eqref{conmom}, \eqref{covmom} and \eqref{squmom} respectively, we can define the momentum uncertainty as
\begin{align}
\sigma_p\equiv\sqrt{\braket{\hat{p}^2}-\braket{\hat{p}^i}\braket{\hat{p}_i}}.\label{uncdef}
\end{align}
The remainder of this paper will be centered around the evaluation of this quantity.

\subsection{Perturbation around flat space}

Following the approach in \cite{RF2020}, we calculate the uncertainty relation perturbatively. In particular, we expand $G_{ij}$ in Riemann normal coordinates (RNC) $x^i$ to second order
\begin{align}
    G_{ij}&\simeq G^{(0)}_{ij}+G^{(2)}_{ij}\\
    &=\delta_{ij}-\frac{1}{3}R_{ikjl}|_{p_0}x^kx^l
\end{align} 
with the Kronecker Delta $\delta_{ij}$ and the Riemann curvature tensor with respect to the metric $G_{ab}$ evaluated at $p_0.$ This leads to ensuing expansions in $\nabla_i=\nabla_i^{(0)}+\nabla_i^{(2)},\Delta=\Delta_{(0)}+\Delta_{(2)}$ due to
\begin{align}
    \Gamma^i_{jk}\simeq&\left(\Gamma^i_{ij}\right)^{(2)}\\
    =&\frac{1}{3}\left(R^i_{~jkm}+R^i_{~kjm}\right)|_{p_0}x^m.
\end{align}
Additionally, the measure in expanded form reads $\D\mu=\D\mu^{(0)}+\D\mu^{(2)}$, with $\D\mu^{(0)}=\D^3x$ and
\begin{equation}
    \D\mu^{(2)}=-\frac{1}{6}R_{ij}|_{p_0}x^ix^j\D^3x.
\end{equation}
Correspondingly, the scalar product is perturbed as $\braket{}\simeq\braket{}_0+\braket{}_2.$ How to treat quantum mechanical perturbation theory in this special case has been elaborated upon in \cite{RF2020}. Note that with $\braket{}^{(n)}$ we denote the nth correction to the whole amplitude including wave functions and operators inside, while $\braket{}_n$ signifies the nth correction to the scalar product. 

Due to the spherical symmetry of the unperturbed problem, integrals appearing throughout the calculation will be solved in geodesic coordinates $(\sigma,\chi,\gamma)$ which were introduced in \cite{RF2020} and relate to RNC as spherical coordinates to Cartesian ones
\begin{equation}
x^i=\sigma(\sin\chi\cos\gamma,\sin\chi\sin\gamma,\cos\chi).
\end{equation}
As the metric has been expanded in RNC, the same should be done for the shift, which now reads
\begin{align}
    N^i\simeq&N^i_{(0)}+N^i_{(1)}+N^i_{(2)}\\
    =&N^i|_{p_0}+\nabla_jN^i|_{p_0}x^j+\nabla_j\nabla_kN^i|_{p_0}x^jx^k.
\end{align}
This means that, in principle, the shift could yield zeroth and first order corrections. 

Finally, the momentum uncertainty is expanded as
\begin{align}
\sigma_p\simeq&\sqrt{\var{p}^{(0)}+\var{p}^{(1)}+\var{p}^{(2)}}\\
        =&\sigma_p^{(0)}\left(1+\frac{\var{p}^{(1)}}{2\var{p}^{(0)}}+\frac{\var{p}^{(2)}-\frac{\var{p}^{(1)}}{4\var{p}^{(0)}}}{2\var{p}^{(0)}}\right)
\end{align}
where we introduced the variance $\sigma_p^2.$ These contributions will be treated order by order.

\subsection{Unperturbed uncertainty relation}

To zeroth order the momentum uncertainty reads
\begin{equation}
    \sigma_p^{(0)}=\sqrt{\var{\pi}^{(0)}+\frac{2\hbar}{\lambdabar_C}N^i|_{p_0}\braket{i\hbar\partial_i+\pi_i}^{(0)}+\frac{\hbar^2}{\lambdabar^2_C}\var{N}^{(0)}}\label{0thorderunc}
\end{equation}
where we used that $i\partial_i$ is hermitian with respect to the unperturbed measure.

As $\pi_i^{(0)}=-i\hbar\partial_i,$ the term in the brackets vanishes. Moreover, it is a simple exercise to show that a similar cancellation occurs to the variance of the shift leaving us with
\begin{equation}
    \sigma_p^{(0)}=\sigma_\pi^{(0)}.
\end{equation}
Hence, the shift has no influence to zeroth order.

Note that these considerations hold independently of the state with respect to which the uncertainty is calculated. According to the formalism developed in \cite{Schuermann2018} and subsequently in \cite{RF2020}, we obtain an uncertainty relation by finding the state of minimum momentum uncertainty. 

Yet, there it was falsely claimed that $\braket{\hat{\pi}^i}=\braket{\hat{\pi}_i}=0$ for arbitrary states in arbitrary backgrounds. This is indeed not the case. Nevertheless, as shown in the appendix, the state of minimal uncertainty is still the ground state for flat backgrounds. In geodesic coordinates, it reads
\begin{equation}
    \psi^{(0)}_{100}=\frac{1}{\sqrt{2\pi\rho}}\frac{\sin\left(\pi\frac{\sigma}{\rho}\right)}{\sigma}\label{groundstate}
\end{equation}
which is the leading-order contribution to $\psi_{100}.$ Consequently, the relation
\begin{equation}
    \sigma_p^{(0)}\left(\Psi^{(0)}\right)\geq\sigma_p^{(0)}\left(\psi_{100}^{(0)}\right)
\end{equation}
holds for the zeroth-order approximation $\Psi^{(0)}$ to every $\Psi\in\hil.$ As perturbations are assumed to be small, they cannot change this relation, which is why we deduce that
\begin{equation}
    \sigma_p(\Psi)\geq\sigma_p\left(\psi_{100}\right).
\end{equation}
As shown in the appendix, the ground state (as all eigenstates of the Laplace-Beltrami operator) satisfies
\begin{equation}
    \braket{\psi_{100}|\hat{\pi}^i\psi_{100}}=\braket{\psi_{100}|\hat{\pi}_i\psi_{100}}=0\label{expconjmomgeod}
\end{equation}
to all orders. Therefore, we obtain
\begin{equation}
    \sigma^{(0)}_p\geq\sqrt{-\hbar^2\braket{\psi_{100}|\Delta\psi_{100}}^{(0)}}=\hbar\pi/\rho.
\end{equation}
Now that the state of lowest momentum uncertainty is identified, it is time to evaluate the curvature induced corrections.

\subsection{Corrections}

First, according to \eqref{pmix}, we can generally write
\begin{equation}
\braket{\hat{p}_{mix}^2}=\hbar\text{Im}\braket{N^i\partial_i}.
\end{equation}
As the ground state is real, the integrand appearing in
\begin{align}
    \braket{\psi_{100}|N^i\partial_i\psi_{100}}=\int\D\mu \psi_{100} N^i\partial_i\psi_{100}
\end{align}
is purely real and so is the integral. Thus, in the ground state the expectation value of $\hat{p}^2_{mix}$ vanishes.

Furthermore, terms mixing expectation values of the shift and the momentum vanish identically due to \eqref{expconjmomgeod}. Then, the variance of the ground state equals
\begin{equation}
    \sigma_p^2\left(\psi_{100}\right)=\sigma_{\pi}^2\left(\psi_{100}\right)+\frac{\hbar^2}{\lambdabar_C^2}\sigma_{N}^2\left(\psi_{100}\right)
\end{equation}
non-perturbatively. Curvature corrections to $\var{\pi}$ appear to second order and the first-order correction to $\var{N}$ is subject to similar cancellations as in \eqref{0thorderunc}. Hence, the first-order contribution to the variance of the physical momentum operator vanishes
\begin{equation}
    \var{p}^{(1)}(\psi_{100})=0
\end{equation}
from which we deduce that the shift corrects the uncertainty relation at the same order as the background curvature.

Moreover, from \cite{RF2020} we already know that
\begin{equation}
    \var{\pi}^{(2)}(\psi_{100})=-\frac{1}{6}R|_{p_0}
\end{equation}
where $R|_{p_0}=G^{ij}G^{kl}R_{ikjl}|_{p_0}.$ At this point, we are left with the second-order correction to the variance, which after cancellations reads
\begin{align}
    \var{N}^{(2)}=\nabla_k N^i\nabla_l N^j G_{ij}|_{p_0}\left(\braket{x^k x^l}-\braket{x^k}\braket{x^j}\right)^{(0)}.
\end{align}
When evaluated with respect to the ground state, the second term in the bracket vanishes while the first yields
\begin{equation}
    \braket{\psi_{100}|x^jx^k\psi_{100}}^{(0)}=\frac{\rho^2}{18}\left(2-\frac{3}{\pi^2}\right)\delta^{jk}.
\end{equation}
Finally, lowering and raising indices with the effective metric $G_{ij},$ the second-order correction to the variance of the physical momentum operator equals
\begin{equation}
    \var{p}^{(2)}(\psi_{100})=-\frac{R|_{p_0}}{6}+\xi\frac{\rho^2}{2\lambdabar_C^2}\nabla_jN_i\nabla^jN^i|_{p_0}
\end{equation}
where we introduced the mathematical constant $\xi=(2-3/\pi^2)/9.$

Observe that, though it seems to be of higher order at first glance due to the factor $\rho^2,$ the second term is actually of second order because the expansion done here is performed in terms of $\rho\sqrt{\mathcal{R}}$ where $\mathcal{R}$ denotes any curvature invariant with dimensions of squared inverse length. Thus, no assumptions were made concerning the factor $\rho^2/\lambdabar^2_C.$

\subsection{Result}

Gathering all the results from the previous sections and introducing the Compton wavelength $\lambda_C\equiv2\pi\lambdabar_C$, we obtain the uncertainty relation:
\begin{align}
    \sigma_p\rho\gtrsim &\pi\hbar\left[1-\frac{\rho^2R|_{p_0}}{12\pi^2}+\xi\frac{\rho^4}{\lambda_C^2}\nabla_jN_i\nabla^jN^i|_{p_0}\right].\label{finalunc}
\end{align}
In short, given a four metric $g_{\mu\nu}$ and an observer defining a foliation of space-time, the uncertainty relation \eqref{finalunc} can be computed by evaluating the Ricci scalar derived from $G_{ij}$ and the corresponding covariant derivative of the shift at $p_0.$ 

\section{Applications\label{sec:app}}

In what follows, we will employ the formalism developed up until now to present explicit results related to several relevant metrics. To achieve non-trivial solutions, we shall require to work with space-time metrics for which the shift vector $N^i$ in the $3+1$ decomposition is non-vanishing, otherwise the physical momentum \eqref{phys} coincides with the conjugate momentum $\pi_i.$ For the sake of continuity with Ref. \cite{RF2020}, we start our analysis with the G\"odel universe. Subsequently, we focus our attention on the first weak-field solution for a rotating source in the context of General Relativity (GR), namely the Lense-Thirring space-time. As the last two examples, we investigate space-times stemming from rotating compact objects in the framework of extended theories of gravity with the purpose of pinpointing the main differences with respect to the standard GR scenario.

\subsection{G\"odel universe}

The G\"odel solution \cite{godel} is a homogeneous and anisotropic space-time arising from Einstein's field equations for a perfect fluid with non-vanishing angular momentum. It essentially describes a rotating universe in which closed timelike curves are allowed, thus in principle permitting time travel. The line element associated with such a curved background written in cylindrical coordinates $(t,r^a)=(t,r,\phi,z)$ reads \cite{godel}
    \begin{align}\nonumber
\D s^2  =&-\D t^2-\frac{2r^2}{a\sqrt{2}}\D t \D\phi+\frac{\D r^2}{\left(1+\frac{r^2}{4a^2}\right)}\\
        &+r^2\left(1-\frac{r^2}{4a^2}\right)\D\phi^2+\D z^2\label{eqgodel}
\end{align}
with the constant parameter $a>0$ which has units of length and quantifies the angular momentum of matter. Have in mind that this slicing only covers the region $r< 2a.$

An observer orbiting circularly around the $z$-axis (\ie co-rotating with the Gödel universe) will experience the flow of proper time according to the time coordinate $t.$ The effective lapse, three-metric and shift from the point of view of this observer read
\begin{align}
N                                       =&\sqrt{1+\frac{2}{4\frac{a^2}{r^2}-1}}\\
G_{ab}\D r^a\D r^b                      =&\frac{1}{N}\left[\frac{\D r^2}{1+\frac{r^2}{4a^2}}+\left(1-\frac{r^2}{4a^2}\right)r^2\D\phi^2+\D z^2\right]\\
N^a\frac{\partial}{\partial r^a}        =&-\frac{a}{\sqrt{2}\left(a^2-\frac{r^2}{4}\right)}\partial_\phi.
\end{align}
As $r<2a$ in this slicing and the prefactors of terms containing higher powers of $r_0/2a$ (where $r_0=r|_{p_0}$) in the resulting uncertainty relation get ever smaller, we will only display the next-to-leading order for the sake of brevity.

Correspondingly, the observer defined above measures the uncertainty relation
\begin{equation}
    \sigma_p\rho\gtrsim \pi\hbar\left[1-\frac{\rho^2}{4\pi^2a^2}\left(1-\xi \frac{\rho^2}{\lambdabar_C^2}+\frac{17}{24}\frac{r_0^2}{a^2}\right)\right].
\end{equation}

\subsection{Lense-Thirring solution\label{subsec:lense}}

The phenomenon of frame-dragging was discovered only few years after the final settlement of GR. As a matter of fact, in 1918 Lense and Thirring found the weak-field limit for the space-time generated by a rotating body \cite{lense}. The main prediction of this solution is the existence of a precession of the orbits drawn by a test body, a feature that is completely absent in Newtonian mechanics. To get to this conclusion, they argued that in isotropic spherical coordinates $(t,r^e)=(t,r,\theta,\varphi)$ the metric tensor originating from a rotating source takes the form \cite{lense}
\begin{align}\label{eqlense}
\D s^2= &-(1+2\phi_{GR})\D t^2+(1-2\phi_{GR})\left(\D r^2+r^2\D\Omega^2\right)\nonumber\\
        &+4\phi_{GR} a_J\sin^2\theta \D\varphi \D t
\end{align}
where $\phi_{GR}=-GM/r$ is the usual Newtonian potential whereas $a_J=J/M$ is the rotational parameter with $J$ denoting the angular momentum of the source. 

Note that the time coordinate $t$ used here corresponds to the time measured by a static observer at infinite distance from the gravitating body in the center. In turn, the uncertainty is calculated as it would be measured by this observer. In the non-relativistic limit, though, this provides a good approximation of the slicing carved out by the dynamical rest frame of the particle itself. Therefore, corrections to the results are expected to be of higher order. Analogous considerations apply to the extended models of gravity which are treated as corrections to \eqref{eqlense} below.

The static observer at infinity experiences the effective metric, shift and lapse
\begin{align}
N                                           \simeq&1-\phi_{GR}\\
N^e\frac{\partial}{\partial r^e}            \simeq&-2a_J\frac{\phi_{GR}}{r^2}\\
    G_{ef}\D r^e\D r^f                      \simeq&\left(1+3\phi_{GR}\right)\left(\D r^2+r^2\D\Omega^2\right).
\end{align}
Unfortunately, this reasoning leads to an uncertainty which is at least quadratic in the gravitational potential while the Lense-Thirring solution corresponds to a first-order expansion. In light of this, we generalize the discussion by approximating the Kerr metric in Boyer-Lindquist coordinates
\begin{align}
    \D s^2=&-\left(1+2\tilde{\phi}_{GR}\frac{\tilde{r}^2}{\Xi^2}\right)\D t^2+4\tilde{\phi}_{GR}a_J\sin^2\theta\frac{\tilde{r}^2}{\Xi^2}\D t\D \varphi\nonumber\\
    &+\frac{\Xi^2}{\Sigma}\D \tilde{r}^2+\Xi^2\D\theta^2+\tilde{r}^2\Big(1+\frac{a_J^2}{\tilde{r}^2}\nonumber\\
    &-2\tilde{\phi}_{GR}\sin^2\theta\frac{a_J^2}{\Xi^2}\Big)\sin^2\theta\D\varphi^2
\end{align}
with $\tilde{\phi}_{GR}=-GM/\tilde{r}$, $\Xi=\tilde{r}\sqrt{1+\cos^2\theta a_J^2/\tilde{r}^2}$ and $\Sigma=\tilde{r}^2(1+\tilde{\phi}_{GR}+a_J^2/\tilde{r}^2)$ to fourth order in $\tilde{\phi}_{GR}$ and $a_J/\tilde{r}$ simultaneously. Bear in mind that the Schwarzschild-like $\tilde{r}$ relates to the radial coordinate introduced with the Lense-Thirring metric as $\tilde{r}=r(1-\phi_{GR}/2)^2.$ As the resulting uncertainty relation is 3-diffeomorphism invariant, we will nevertheless provide it in terms of $r$ for the sake of future convenience.

Consequently, the static observer at infinity measures the uncertainty relation
\begin{align}
    \sigma_p\rho\geq&\pi\hbar\Bigg\{1+\phi_{GR}^2\frac{\rho^2}{48\pi^2r^2}\Bigg[10+30\phi_{GR}+55\phi_{GR}^2-\frac{a_J^2}{r^2}\nonumber\\
    &\times\Big(469-217\cos 2\theta-96\xi \frac{\rho^2}{\lambdabar_C^2}\left(7-3\cos 2\theta\right)\Big)\Bigg]\Bigg\}\Bigg|_{p_0}\\
    \equiv&\pi\hbar\left(1+\lambda_{LT}\right)
\end{align}
where the last line defines the (generalized) Lense-Thirring correction $\lambda_{LT}.$ We would like to remark that the dimensionless number $\rho^2/\lambdabar_C^2$ is unconstrained; for this reason, in a suitable regime it can increase the effect of space-time rotation.

\begin{figure}
    \centering
    \includegraphics[width=\linewidth]{"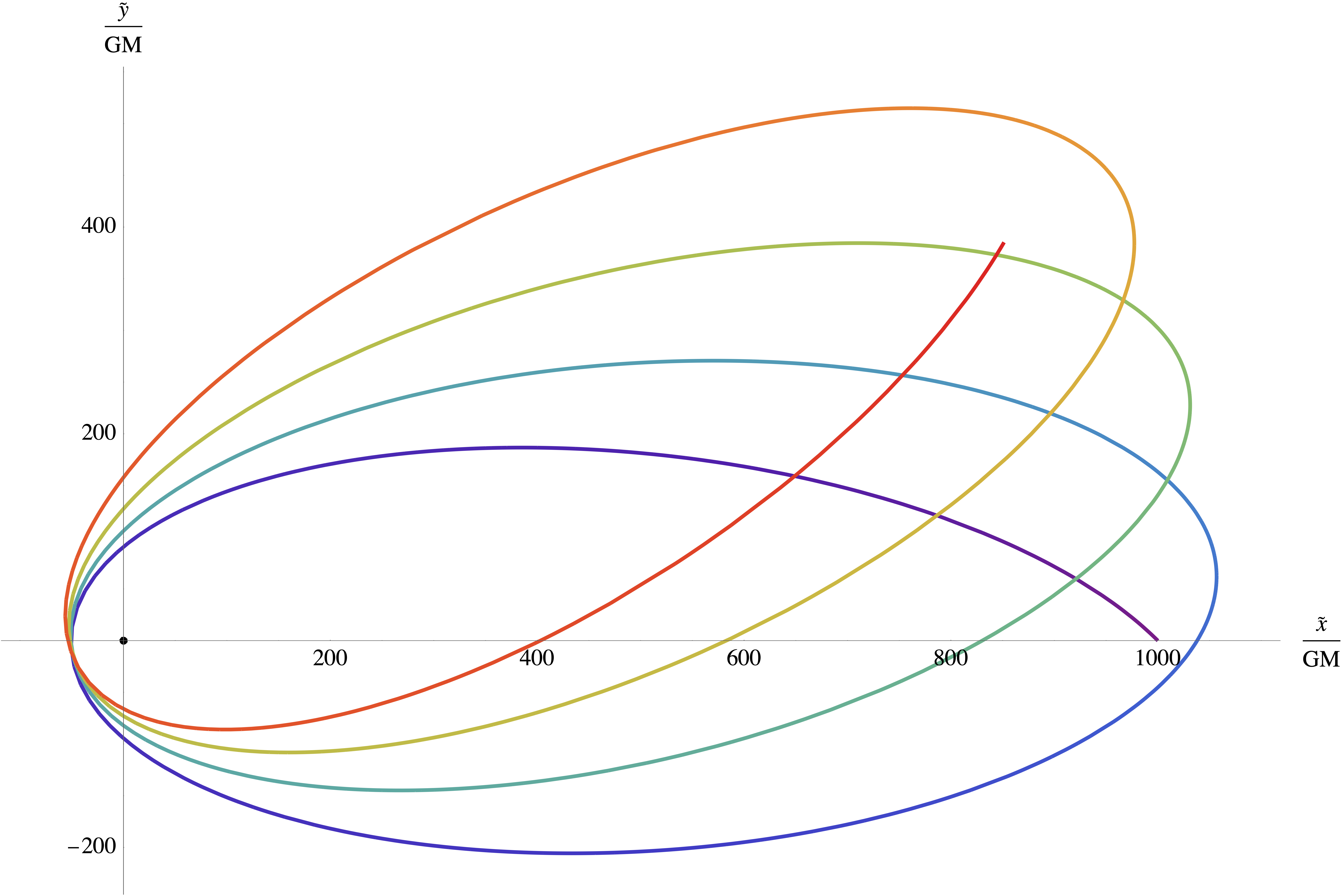"}
    \caption{The trajectory followed by a massive particle in the equatorial ($\tilde{x}$-$\tilde{y}$-) plane of the expanded Kerr metric as seen from above for $a_J/GM=0.5$. The starting point lies on the $x/GM$ axis at a distance $1000$ from the source with initial velocity (in coordinates $(t,r^e)$) $u(\tau =0)=(1.001,-0.010,0.000,0.000).$ The color ranging from violet to red indicates an increase in proper time $\tau$ whereas the black disk at the center symbolizes the outer horizon.}
    \label{fig:orb}
\end{figure}

\begin{figure}
    \centering
    \includegraphics[width=\linewidth]{"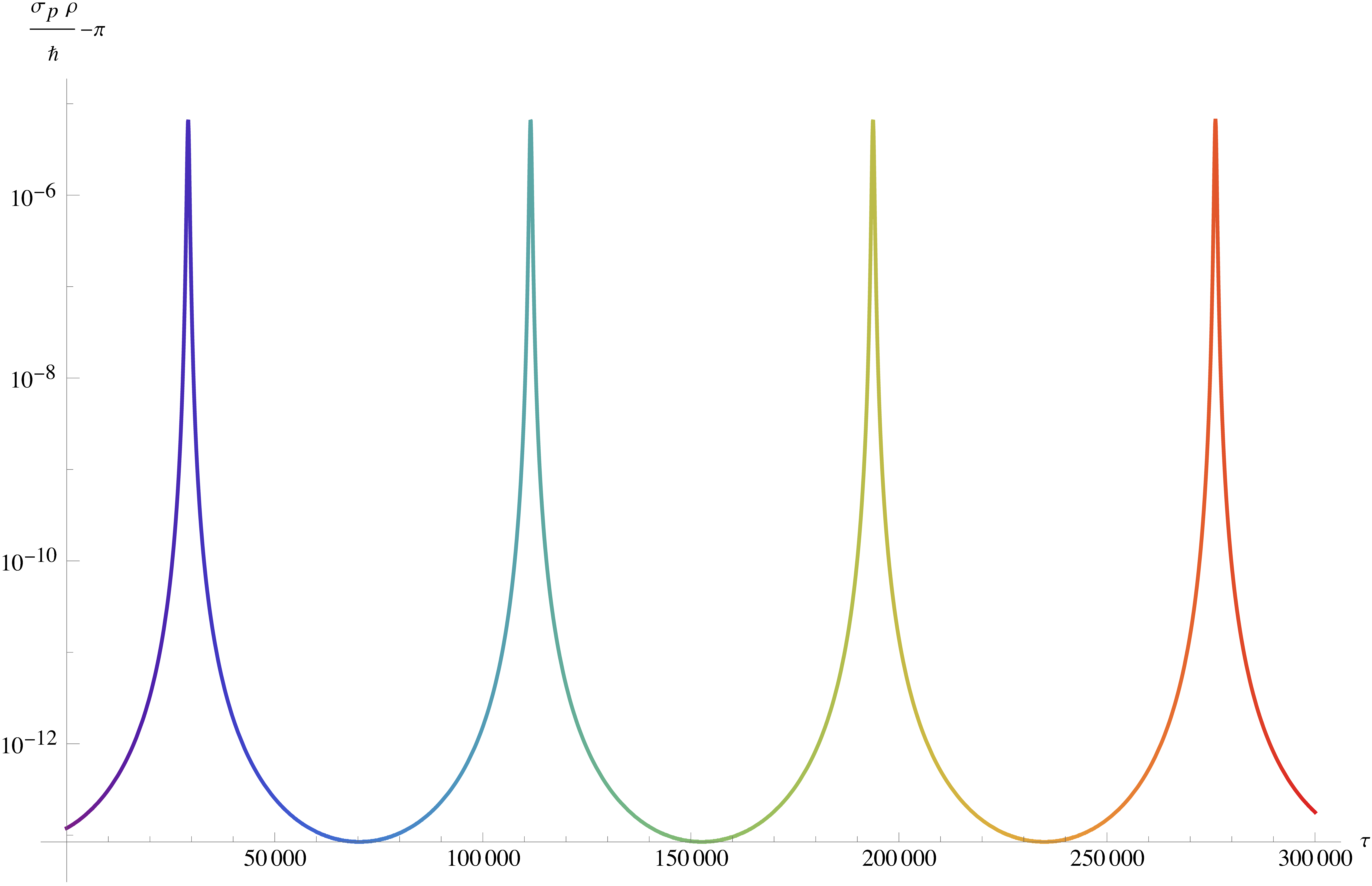"}
    \caption{The corrections to the uncertainty relation experienced along the trajectory in figure \ref{fig:orb} by a particle of Compton wavelength $\lambda_C=10^{-4}GM$ as a function of proper time.}
    \label{fig:uncorb}
\end{figure}

A sample orbit in the equatorial plane of the Kerr metric approximated as indicated above is given in figure \ref{fig:orb}, where the color of the curve changes with increasing proper time. The ensuing correction to the uncertainty relation is displayed in figure \ref{fig:uncorb} as a function of said proper time. The peaks correspond to the positions of smallest distance from the outer horizon along the trajectory. Their height is amplified by the ad-hoc choice of Compton wavelength of the particle $\lambda_C$ and the proximity of the orbit to the horizon which can only be realized surrounding black holes. A more conservative estimate of a measurement done on a geostationary satellite at $r|_{p_0}\sim 10^4\text{km}$ ($GM\sim 10^{-4}\text{km}$) using the largest quantum systems realized presently ($\rho\sim 10^{-15}\text{km}$) leads to a correction of the order $10^{-60}.$

\subsection{Fourth-order gravity}

Fourth-order gravity introduced by Stelle represents one of the first attempts to cure the quantization problems of the gravitational interaction. In particular, it was pointed out \cite{stelle} that the introduction of higher-derivative terms in the Einstein-Hilbert action can make the model renormalizable. To be more precise, according to the prescription in \cite{stelle}, the gravitational action from which to build up quantum gravity should be given by
\begin{equation}\label{actionstelle}
S=\frac{1}{16\pi^2G^2}\int d^4x\sqrt{-g}\left(R+\frac{\alpha}{2\hbar^2}R^2-\frac{\beta}{2\hbar^2}R_{\mu\nu}R^{\mu\nu}\right),
\end{equation}
where $\alpha$ and $\beta$ are dimensionful constants measured in units of inverse mass squared. However, the drawback of this model consists in the appearance of ghost-like degrees of freedom which undermine the unitarity of the underlying quantum field theory. Such a circumstance is a typical feature of local higher-order derivative gravity \cite{shapiro}.

For the current model, a Lense-Thirring-like solution has been recently obtained when analyzing the light bending due to quadratic theories of gravity \cite{luca}. In isotropic spherical coordinates, the aforementioned solution is
\begin{align}\label{geq}
\D s^2= &-(1+2\phi)\D t^2+(1-2\psi)\left(\D r^2+r^2\D\Omega^2\right)\nonumber\\
        &+2\xi\sin^2\theta \D\varphi \D t\,,
\end{align}
where the gravitational potentials are given by
\begin{align}
\phi=& \,\phi_{GR}\left(1+\frac{1}{3}e^{-m_0r/\hbar}-\frac{4}{3}e^{-m_2r/\hbar}\right)\label{eqstelle1}\\[2mm]
\psi=& \,\phi_{GR}\left(1-\frac{1}{3}e^{-m_0r/\hbar}-\frac{2}{3}e^{-m_2r/\hbar}\right)\label{eqstelle2}\\[2mm]
\xi=& \,2\phi_{GR}a_J\left[1-(1+m_2r/\hbar)e^{-m_2r/\hbar}\right]\label{eqstelle3}
\end{align}
with $m_0=2/\sqrt{12\alpha-\beta}$ and $m_2=\sqrt{2/\beta}$ being the masses of the spin-0 and spin-2 massive modes, respectively.

As the influence stemming from the higher-derivative terms ought to be small in comparison to the general relativistic effect, results should be given as corrections to the Lense-Thirring outcome. Written this way, the uncertainty relation reads for small gravitational potentials
\begin{align}
    \sigma_p\rho\geq&\pi\hbar\Bigg[1+\lambda_{LT}\nonumber\\
    &-\frac{\phi_{GR}|_{p_0}}{18\pi^2}\left(\frac{\rho^2}{\lambda_{m_0}^2}e^{-\frac{m_0 r_0}{\hbar}}+\frac{8\rho^2}{\lambda_{m_2}^2}e^{-\frac{m_2 r_0}{\hbar}}\right)\Bigg]
\end{align}
where $\lambda_{m_0}$ and $\lambda_{m_2}$ denote the Compton wavelengths corresponding to the respective massive gravitational modes.

As straightforwardly recognizable in the previous equation, for the current example we do not have to resort to a higher-order expansion of the Kerr-like solution in the context of the examined extended model of gravity, as the leading-order correction is already linear in $\phi_{GR}.$ This feature is shared by the upcoming analysis as well.

\subsection{Infinite-derivative gravity}

Starting from the above scenario and recalling that the reasoning in \cite{shapiro} prevents any local higher-order derivative gravity from being free from ghost fields, it is clear that one must give up on locality to arrive at a quantum gravitational model which is simultaneously renormalizable and unitary. Hovewer, non-locality should be manifest only in the currently unexplored UV regime, since all the available data acquired from gravitational experiments comply with the local behavior of gravity. Along this direction, it is possible to encounter the so-called infinite-derivative gravity theory, which precisely possesses the characteristics listed above. As the name suggests, the usual Einstein-Hilbert action is now accompanied by non-local functions of the curvature invariants; in the simplest form, the non-local gravitational action reads \cite{idg}
\begin{align}\nonumber
S=&\frac{1}{16\pi^2G^2}\int \D^4x\sqrt{-g}\left(R+R\frac{1-e^{-{\hbar^2\Box}/{\kappa^2}}}{4\Box}R\right.\\
&-\left.R_{\mu\nu}\frac{1-e^{-{\hbar^2\Box}/{\kappa^2}}}{2\Box}R^{\mu\nu}\right),\label{actionidg}
\end{align}
where $\kappa$ is the energy scale at which the non-local aspects of gravity are expected to be prominent.

As for the previous example, a Lense-Thirring-like solution can be analytically computed in this framework. Formally, the shape of the metric tensor is the same as the one exhibited in \eqref{geq}, with the difference that here the gravitational potentials are instead represented by
\begin{align}
\phi=\psi=& \,\phi_{GR}\mathrm{Erf}\left(\frac{\kappa\,r}{2\hbar}\right)\label{eqidg1}\\[2mm]
\xi=& \,2\phi_{GR}a_J\left[\mathrm{Erf}\left(\frac{\kappa\,r}{2\hbar}\right)-\frac{\kappa\,r}{\sqrt{\pi}\hbar}e^{-\kappa^2r^2/4\hbar^2}\right]\label{eqidg2}
\end{align}
where $\mathrm{Erf}(x)$ denotes the error function.

Again expressed as corrections to the Lense-Thirring result, the uncertainty relation for small gravitational potentials becomes in this case
\begin{equation}
    \sigma_p\rho\geq\pi\hbar\left(1+\lambda_{LT}-\frac{\phi_{GR}|_{p_0}}{4\pi^{\frac{5}{2}}}\rho^2r_0\frac{\kappa^3}{\hbar^3}e^{-\kappa^2 r_0^2/4\hbar}\right).
\end{equation}

\section{Discussion\label{sec:disc}}

Along the lines of the EUP prescription and by resorting to the dynamics of a non-relativistic particle in a curved background, we have shown how to generally derive a modification of the canonical uncertainty principle surged by the underlying gravitational field in which the motion takes place. To this aim, we have based our considerations on the notion of geodesic ball \cite{Schuermann2009,Schuermann2018} as a 3D-diffeomorphism invariant domain (whose radius yields a measure of position uncertainty). Moreover, we have properly defined a hermitian momentum operator that complies with the canonical commutation relations in the non-relativistic limit of the 3+1 formalism. Finally, perturbing around flat space-time and relying on a result for flat spaces obtained in an earlier work \cite{RF2020}, we have evaluated the gravitationally induced corrections to the uncertainty relation summarized by equation \eqref{finalunc}. To second order, it contains two new contributions, one proportional to the Ricci scalar of the effective three-dimensional metric and one to the squared covariant derivative of the shift vector, both of which are evaluated at the center of the geodesic ball (\ie the expectation value of the position operator).

Furthermore, we have explicitly computed the form of the above uncertainty relation for the G\"odel universe, the Lense-Thirring solution and its extension in the framework of fourth-order and infinite-derivative gravity. Remarkably, whilst the leading-order contribution goes like $\phi_{GR}^2$ in the Lense-Thirring scenario, for the extended models we observe a proportionality to $\phi_{GR}$. Therefore, there may be a regime in which the two terms are comparably important, thus leading to a simultaneous ``coexistence'' of the two quantities. A similar occurrence has also been addressed in different contexts, as for the case of the Casimir effect \cite{casimir}.  

The derivation performed in this paper allows for a dual interpretation of curved energy-momentum spaces replacing geodesic balls in position space with their counterparts in momentum space and the momentum operator with the position operator. Hence, it provides a direct geometrical link between the resulting Generalized Uncertainty Principle and curvature in energy-momentum space as has been anticipated \eg in the geometrical description of Doubly Special Relativity as a theory of de Sitter momentum space. Thus, it paves the way towards momentum-space curvature corrected quantum mechanics on purely geometrical grounds. 

\appendix

\section{Smallest momentum uncertainty state in flat space \label{app: momunc}}

In Refs. \cite{RF2019} and \cite{RF2020}, starting from Schürmann's ideas \cite{Schuermann2018}, the authors wrongly stated that the expectation value of the conjugate momentum operator $\pi_i$ necessarily vanishes due to the Dirichlet boundary conditions. This is in fact the case for general real wave functions $\Psi: {\rm I\!R}^3\rightarrow {\rm I\!R}$ as can be readily verified by
\begin{align}\nonumber
    \braket{\Psi|\hat{\pi}_i\Psi}=&\int\D\mu \Psi\hat{\pi}_i\Psi=-\int\D\mu \hat{\pi}_i(\Psi)\Psi\\
    =&-\braket{\Psi|\hat{\pi}_i\Psi}=0
\end{align}
where we used $[\hat{\pi}_i,\sqrt{\hat{g}}]=0$ and the boundary conditions \eqref{evprob}. An equivalent relation can be derived for the expectation value of the covariant conjugate momentum operator $\braket{\hat{\pi}^i}.$

Yet, the colinearity of the real and imaginary part of the eigenvalue problem \eqref{evprob} does not necessarily imply that the corresponding wave functions are real neither that their momentum expectation values vanish.

A general state $|\Psi\rangle\in\hil$ can be written as
\begin{align}
    |\Psi\rangle&=\sum_{n,l,m}\alpha_{nlm}|\psi_{nlm}\rangle\label{genstate}
\end{align}
with
\begin{align}
\sum_{n,l,m}|\alpha_{nlm}|^2=1\label{coeflim}
\end{align} 
which leads to the expectation values
\begin{align}
    \braket{\Psi|\hat{\pi}_i\Psi}&=\sum_{\substack{nlm\\n'l'm'}}\alpha_{nlm}\alpha_{n'l'm'}^*\braket{\psi_{nlm}|\hat{\pi}_i\psi_{n'l'm'}}\\
    \braket{\Psi|\hat{\pi}^2\Psi}&=\hbar^2\sum_{nlm}|\alpha_{nlm}|^2\lambda_{nl}\label{vanishexprealpsi}
\end{align}
where we used \eqref{flatev} and that $\psi_{nlm}$ are the eigenstates of $\hat{\pi}^2$. 

According to the definition \eqref{uncdef}, the momentum uncertainty can be decreased by additional contributions to the expectation value of the co- and contravariant momentum operators. However, this is accompanied by an increase in the contribution coming from the expectation value of the squared momentum operator as by \eqref{flatev} (assuming a flat background), which more than compensates this.

In this appendix, we will argue that the ground state $\psi_{100}$ continues to yield the smallest conjugated momentum uncertainty $\sigma_\pi=\sqrt{\braket{\hat{\pi}^2}-\braket{\hat{\pi}^i}\braket{\hat{\pi}_i}}$ in flat space, thus saving the results in \cite{RF2020} because the perturbation (naturally being regarded as small) cannot change this choice. A similar argument may also be carried out for spaces of constant curvature as in \cite{Schuermann2018}.

The conjugated momentum operators in flat space simply read $\pi_i=-i\hbar\partial_i$ and $\pi^i=-i\hbar\delta^{ij}\partial_j.$ Obviously, in this case both operators are equivalent. As momentum operators are vector operators, their expectation values are vectors. The choice of z-axis along which they are constructed is arbitrary, which is why the imaginary part of eigenstates of the spherical Laplacian can be rotated away without exception while the expectation value of the momentum operator is changed by an orthogonal transformation. Yet, according to \eqref{vanishexprealpsi} the expectation value of the momentum operator vanishes for real wavefunctions. This implies that the unrotated expectation value has to vanish as well. We conclude that
\begin{equation}
    \braket{\psi_{nlm}|\pi_i\psi_{nlm}}=0.
\end{equation}
Thus, the only non-vanishing contribution to the expectation value of the momentum operator lies in linear combinations of eigenstates of the Laplacian with relative phases. Due to the limitation of the coefficients \eqref{coeflim} and since transition amplitudes generally go like $||\braket{\psi_{nlm}|\pi_i\psi_{n'l'm'}}||\sim\delta_{l,l'\pm 1}$ which makes it impossible to obtain \eg 3 contributions from linear combinations of 3 states, such a combination of any number of basis states in \eqref{genstate} will show the same behaviour as a linear combination of just two of them. Therefore, no further decrease of the momentum uncertainty can be achieved by combining more than two states, which is why we will only deal with this case. Up to a global phase, such a state generally reads
\begin{equation}
    \Phi=\sqrt{a} \psi_{nlm}+e^{i\phi}\sqrt{1-a}\psi_{n'l'm'}\label{lincomb2}
\end{equation}
with the real coefficient $a\in (0,1)$ and the relative phase $\phi\in [0,2\pi).$
For the above state, we obtain the momentum uncertainty 

\begin{align}
    \sigma_\pi(\Phi)  =&\Big[\hbar^2\frac{aj_{l,n}^2+(1-a)j_{l',n'}^2}{\rho^2}\nonumber\\
    &-4a(1-a)||\text{Re}\left(e^{i\phi}\braket{\psi_{nlm}|\hat{\pi}_i\psi_{n'l'm'}}\right)||^2\Big]\\
    \geq & \Big[\hbar^2\frac{aj_{l,n}^2+(1-a)j_{l',n'}^2}{\rho^2}\nonumber\\
    &-4a(1-a)||\text{Abs}\braket{\psi_{nlm}|\hat{\pi}_i\psi_{n'l'm'}}||^2\Big].
\end{align}

In order for the resulting state to be the one of smallest uncertainty, it has to satisfy $\sigma_\pi(\psi_{100})>\sigma_\pi(\Phi)$ which can be recast as
\begin{equation}
    ||\text{Abs}\braket{\psi_{nlm}|\hat{\pi}_i\psi_{n'l'm'}}||^2\frac{\rho^2}{\hbar^2}>C(a)^2
\end{equation}
where we introduced the function
\begin{equation}
    C(a)=\sqrt{\frac{j_{l',n'}^2+\frac{a}{1-a}j_{n',l'}^2-\frac{\pi^2}{1-a}}{4a}}.
\end{equation}
As for the allowed $n,l$ we have $\pi\leq j_{n,l};$ this function diverges positively for $a\rightarrow 0$ and $a\rightarrow 1$ and is continuous in between. Hence, it has to reach a minimum for $0\leq a_{min}\leq 1$ such that $C(a)\geq C(a_{min})$ at which it reads
\begin{widetext}
\begin{equation}
    C(a_{min})=\sqrt{\frac{(j_{n,l}^2-j_{n',l'}^2)^2\sqrt{(j_{n,l}^2-\pi^2)(j_{n',l'}^2-\pi^2)}}{\left[\pi^2-j_{n,l}^2+\sqrt{(j_{n,l}^2-\pi^2)(j_{n',l'}^2-\pi^2)}\right]\left[\pi^2-j_{n',l'}^2+\sqrt{(j_{n,l}^2-\pi^2)(j_{n',l'}^2-\pi^2)}\right]}}.
\end{equation}
\end{widetext}
We now see that the transition amplitude of any linear combination of two eigenstates of the Laplace-Beltrami operator whose uncertainty is smaller than the one of the ground state has to satisfy the inequality
\begin{equation}
    ||\text{Abs}\braket{\psi_{nlm}|\hat{\pi}_i\psi_{n'l'm'}}||\frac{\rho}{\hbar}>C(a_{min})\label{ineq}
\end{equation}
which is independent of the parameters $a$ and $\phi.$

\begin{figure}
    \centering
    \includegraphics[width=\linewidth]{"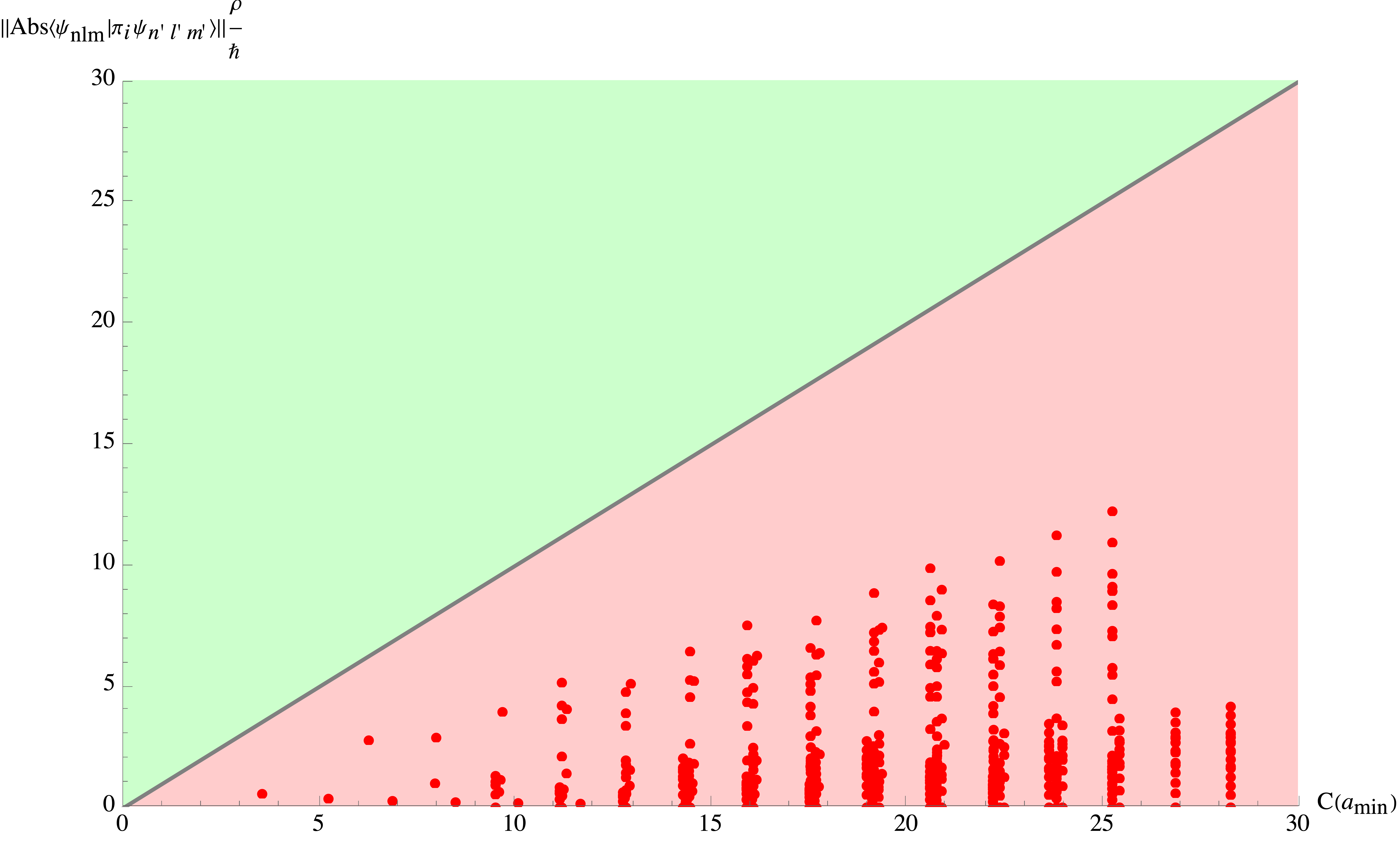"}
    \caption{Visualisation of numerical results for the left-hand side and the right-hand side of \eqref{ineq} for linear combinations of all eigenstates of the Laplacian with $n\leq 6,$ $n'\leq 7$. Numerical results are symbolized by red dots. The inequality is satisfied in the green area while it does not hold in the red area.}
    \label{fig:ineq}
\end{figure}

The right-hand and left-hand sides of this inequality are displayed in figure \ref{fig:ineq} for all eigenstates of the Laplacian with $n\leq 6$ and $n'\leq 7$ where the green area stands for results that would satisfy it. Obviously, no state satisfies \eqref{ineq} and the difference to the line separating the two generally increases with increasing $n,l,n',l'.$

There is no reason to expect this to change for higher values of $n,l,n',l'.$ Therefore, we conclude that the ground state $\psi_{100}$ is indeed the state of lowest uncertainty in flat space.

\section*{Acknowledgments}

The work of F.W. was supported by the Polish National Research and Development Center (NCBR) project ''UNIWERSYTET 2.0. --  STREFA KARIERY'', POWR.03.05.00-00-Z064/17-00 (2018-2022).


\begin{thebibliography}{99}

\bibitem{Amati87} D.~Amati, M.~Ciafaloni and G.~Veneziano,  Phys. Lett. B {\bf 197}, 81 (1987).

\bibitem{Kempf95} A.~Kempf, G.~Mangano and R.B.~Mann, Phys. Rev. D {\bf 52}, 1108 (1995).

\bibitem{Maggiore93} M.~Maggiore, Phys. Lett. B {\bf 304}, 65 (1993).

\bibitem{Adler01} R.J.~Adler, P.~Chen, and D.I.~Santiago, Gen. Rel. Grav. {\bf 33}, 2101 (2001). 

\bibitem{GUPReview} S.~Hossenfelder, Living Rev. Relativity {\bf 16}, 2 (2013). 

\bibitem{AdlerDuality} R.J.~Adler and D.I.~Santiago, Mod. Phys. Lett. A {\bf 14}, 1371 (1999). 

\bibitem{Sparsity} A.~Alonso-Serrano, M.P.~D\c{a}browski and H.~Gohar, Phys. Rev. D {\bf 97}, 044029 (2018).

\bibitem{noncom}
S.~Capozziello, G.~Lambiase, and G.~Scarpetta, Int. J. Theor. Phys. \textbf{39}, 15 (2000).

\bibitem{noncom2} 
T.~Kanazawa, G.~Lambiase, G.~Vilasi, and A.~Yoshioka, Eur. Phys. J. C \textbf{79}, 95 (2019).

\bibitem{noncom3}
G.G.~Luciano and L.~Petruzziello,
  Eur.\ Phys.\ J.\ C {\bf 79}, 283 (2019).

\bibitem{bis}
R.J.~Adler, P.~Chen, and D.I.~Santiago, Gen. Rel. Grav. \textbf{33}, 2101 (2001).

\bibitem{ter}
F.~Scardigli and R.~Casadio,
Class. Quant. Grav. \textbf{20}, 3915 (2003).

\bibitem{ter2}
P.~Jizba, H.~Kleinert, and F.~Scardigli, Phys. Rev. D \textbf{81}, 084030 (2010).

\bibitem{quar}
P.~Chen, Y.C.~Ong, and Dh.~Yeom, Phys. Rep. \textbf{603}, 1 (2015).

\bibitem{ses}
F.~Scardigli, G.~Lambiase, and E.~Vagenas, Phys. Lett. B \textbf{767}, 242 (2017).

\bibitem{ong}
Y.C.~Ong, JCAP \textbf{1809}, 015 (2018).

\bibitem{set}
L.~Buoninfante, G.G.~Luciano, and L.~Petruzziello,
  Eur.\ Phys.\ J.\ C {\bf 79}, 663 (2019).
  
\bibitem{ot}
  L.~Buoninfante, G.~Lambiase, G.G.~Luciano, and L.~Petruzziello,
  Eur. Phys. J. C \textbf{80}, 853 (2020).
	
\bibitem{nov}	
L.~Buoninfante, G.G.~Luciano, L.~Petruzziello, and F.~Scardigli,
arXiv:2009.12530 [hep-th] (2020).

\bibitem{dieci}
L.~Petruzziello,
arXiv:2010.05896 [hep-th] (2020).

\bibitem{noui} 
  K.~Nouicer,
  J.\ Phys.\ A {\bf 38}, 10027 (2005).

\bibitem{frassino}
A.M.~Frassino and O.~Panella, Phys. Rev. D \textbf{85}, 045030 (2012).

\bibitem{qft}
S.~Hossenfelder,
  Living Rev.\ Rel.\  {\bf 16}, 2 (2013).

\bibitem{qft2}
F.~Scardigli, M.~Blasone, G.~Luciano, and R.~Casadio, Eur. Phys. J. C \textbf{78}, 728 (2018).

\bibitem{qft3}
M.~Blasone, G.~Lambiase, G.G.~Luciano, L.~Petruzziello, and F.~Scardigli,
  J.\ Phys.\ Conf.\ Ser.\  {\bf 1275}, 012024 (2019).

\bibitem{qft4}
M.~Blasone, G.~Lambiase, G.G.~Luciano, L.~Petruzziello, and F.~Scardigli,	
  Int.\ J.\ Mod.\ Phys.\ D {\bf 29}, 2050011 (2020).	

\bibitem{czech}
C.~Quesne and V.~Tkachuk,
J. Phys. A \textbf{39}, 10909-10922 (2006).

\bibitem{czech2}
V.~Todorinov, P.~Bosso, and S.~Das,
Ann. Phys. (N.Y.) \textbf{405}, 92 (2019).

\bibitem{pikovski}
I.~Pikovski, M.R.~Vanner, M.~Aspelmeyer, M.S.~Kim, and C.~Brukner,
Nature Phys. \textbf{8}, 393 (2012).

\bibitem{plenio}
S.P.~Kumar and M.B.~Plenio,
Phys. Rev. A \textbf{97}, 063855 (2018).

\bibitem{plenio2}
S.P.~Kumar and M.B.~Plenio,
Nature Commun. \textbf{11}, 3900 (2020).

\bibitem{dasprl}
S.~Das and E.C.~Vagenas,
Phys. Rev. Lett. \textbf{101}, 221301 (2008).

\bibitem{dasdecoherence}
S.~Das, M.P.G.~Robbins and E.C.~Vagenas,
Int. J. Mod. Phys. D \textbf{27}, 1850008 (2017).

\bibitem{decoherence}
L.~Petruzziello and F.~Illuminati,
arXiv:2011.01255 [gr-qc] (2020).

\bibitem{Mignemi2010} S.~Mignemi, Mod. Phys. Lett. A{\bf 25}, 1697 (2010). 

\bibitem{Anrade} R.N.~Costa~Pilho, J.P.M.~Braga, J.H.S.~Lira and J.S.~Anrade, Phys. Lett. B {\bf 755}, 367 (2016). 

\bibitem{Mureika} J.R.~Mureika, Phys. Lett. B {\bf 789}, 88 (2019). 

\bibitem{EUPThermod} T.~Zhu, J.-R.~Ren and M.-F.~Li, Phys. Lett. B {\bf 674}, 204 (2009).  

\bibitem{Bambi2008} C.~Bambi and F.R.~Urban, Class. Quantum Grav. {\bf 25}, 095006 (2008). 

\bibitem{Schuermann2009} T.~Sch\"urmann and I. Hofmann, Found. Phys. {\bf 39}, 958 (2009). 

\bibitem{Schuermann2018} T.~Sch\"urmann,  Found.\ Phys.\  {\bf 48}, 716 (2018).

\bibitem{RF2019} M.P.~D\c{a}browski and F.~Wagner, Eur. Phys. J. C {\bf 79}, 716 (2019). 

\bibitem{RF2020} M.P.~D\c{a}browski and F.~Wagner, Eur. Phys. J. C {\bf 80}, 676 (2020). 

\bibitem{gravitomagnetism}
I.~Ciufolini and J.A.~Wheeler, \emph{Gravitation and Inertia} (Princeton University Press, New Jersey,1995).

\bibitem{DeWitt:1952js}
B.S.~DeWitt,
Phys. Rev. \textbf{85}, 653 (1952).

\bibitem{ADM} 
R.L.~Arnowitt, S.~Deser and C.W.~Misner,
Phys. Rev. \textbf{116}, 1322 (1959);
R.L.~Arnowitt, S.~Deser and C.W.~Misner,
Phys. Rev. \textbf{117}, 1595 (1960);
R.L.~Arnowitt, S.~Deser and C.W.~Misner,
Gen. Rel. Grav. \textbf{40}, 1997 (2008).

\bibitem{eric}
E.~Gourgoulhon, \emph{3+1 Formalism in General Relativity} (Springer, Berlin/Heidelberg, 2012).

\bibitem{godel}
K.~Godel,
Rev. Mod. Phys. \textbf{21}, 447 (1949);
E.~Kajari, R.~Walser, W.P.~Schleich and A.~Delgado,
Gen. Rel. Grav. \textbf{36}, 2289 (2004);
M.~Buser, E.~Kajari and W.P.~Schleich,
New J. Phys. \textbf{15}, 013063 (2013).

\bibitem{lense}
J.~Lense and H.~Thirring,
Phys. Z. \textbf{19}, 156 (1918).

\bibitem{stelle}
K.S.~Stelle,
Phys. Rev. D \textbf{16}, 953 (1977);
K.S.~Stelle,
Gen. Rel. Grav. \textbf{9}, 353-371 (1978).

\bibitem{shapiro}
M.~Asorey, J.L.~Lopez and I.L.~Shapiro,
Int. J. Mod. Phys. A \textbf{12}, 5711 (1997).

\bibitem{luca}
L.~Buoninfante and B.L.~Giacchini,
Phys. Rev. D \textbf{102}, 024020 (2020).

\bibitem{idg}
T.~Biswas, E.~Gerwick, T.~Koivisto and A.~Mazumdar,
Phys. Rev. Lett. \textbf{108}, 031101 (2012);
L.~Buoninfante, A.S.~Koshelev, G.~Lambiase and A.~Mazumdar,
JCAP \textbf{09}, 034 (2018);
L.~Buoninfante, G.~Lambiase and A.~Mazumdar,
Nucl. Phys. B \textbf{944}, 114646 (2019).

\bibitem{casimir}
L.~Buoninfante, G.~Lambiase, L.~Petruzziello and A.~Stabile,
Eur. Phys. J. C \textbf{79}, 41 (2019).

\end{thebibliography}
\end{document}